\documentclass[11pt]{article}

\usepackage[utf8]{inputenc}
\usepackage[T1]{fontenc}
\usepackage{lmodern}
\usepackage[a4paper,margin=2.5cm]{geometry}
\usepackage{microtype}
\usepackage{amsmath,amssymb}

\DeclareMathSymbol{\Gamma}{\mathalpha}{letters}{"00}
\DeclareMathSymbol{\Phi}{\mathalpha}{letters}{"08}
\DeclareMathSymbol{\Omega}{\mathalpha}{letters}{"0A}
\usepackage{graphicx}
\usepackage{enumitem}
\usepackage[numbers,sort&compress]{natbib}
\usepackage{hyperref}

\hypersetup{hidelinks}

\setlist[itemize]{leftmargin=1.7em,itemsep=0.35em,topsep=0.35em}
\title{Electromagnetic information theory \\for programmable metasurfaces}

\author{Philipp del Hougne$^{1,2,*}$, Marco Di Renzo$^{3,4}$, Owen D. Miller$^{5}$, David A. B. Miller$^{6}$\\[0.8em]
\small $^{1}$CNRS, IETR -- Univ Rennes, France\\
\small $^{2}$Aalto University, Finland\\
\small $^{3}$CNRS, IETR -- CentraleSup\'elec, France\\
\small $^{4}$King's College London, UK\\
\small $^{5}$Yale University, USA\\
\small $^{6}$Stanford University, USA\\[0.5em]
\small $^{*}$philipp.del-hougne@univ-rennes.fr
}

\date{}

\begin{document}

\maketitle

\begin{abstract}
Programmable metasurfaces enable fine-grained control of electromagnetic waves and are creating new opportunities for information extraction, information transfer and information processing. Yet, despite exciting experimental demonstrations, their fundamental capabilities remain poorly understood. In this Perspective, we argue that programmable metasurfaces, and programmable wave systems more generally, call for an electromagnetic information theory that addresses limits at two levels, in a unified manner across applications like communications, sensing and computing. \textit{First,} for a given physical platform, such a theory should bound achievable information manipulations, such as sensing precision, information-transfer rates, and computational performance. \textit{Second,} at a more fundamental level, it should determine what non-programmable and programmable physical resources bound these system-specific limits, and what minimum resources are required to attain a prescribed level of performance. We discuss emerging theoretical approaches to these questions, and their connection to experimentally realizable systems. We examine how device thickness, programmable degrees of freedom, nonlocality, coupling and multiple scattering constrain these limits. Establishing the envisioned framework could clarify the ultimate capabilities of programmable metasurfaces and guide the design of future wave-based information systems.
\end{abstract}

\section{Introduction}
\label{sec:introduction}

Electromagnetic waves, and their interactions with matter, allow us to manipulate information. Imaging and sensing systems extract information carried by impinging fields or encoded in how objects and scenes scatter waves; communication systems rely on electromagnetic waves to transfer information; and emerging analog wave-domain computing schemes exploit tailored wave--matter interactions to implement desired mathematical operations. Beyond demonstrating such capabilities, a fundamental challenge is to understand their ultimate limits: the precision with which a physical system can extract information, the rate at which it can transfer information, and the fidelity and complexity of the mathematical operations that it can implement.

The ability to engineer wave--matter interactions has advanced rapidly. Structured materials and, in particular, metasurfaces have enabled increasingly compact and versatile control of electromagnetic waves. In parallel, a substantial theoretical understanding has emerged of the fundamental capabilities of fixed wave systems. A common thread is that the geometry, material response and spatial extent of a physical system constrain the electromagnetic transformations that it can implement~\cite{Miller2007FundamentalLimit,ehrenborg2020physical,Gustafsson2020UpperBounds,chao2022physical,Miller2026FundamentalLimits}, and thereby its ultimate performance for information extraction, transfer and processing. 

Programmability changes how these fundamental limits must be formulated. For a given information-bearing state, a programmable metasurface provides not only one input--output operator but a family of operators indexed by its control state. If $H(c)$ denotes the input--output operator associated with control variable $c$, and $\mathcal{C}$ the set of physically admissible control states, the relevant object becomes the set of operators $\mathcal{H}=\{H(c):c\in\mathcal{C}\}$, rather than any single optimized operator. While optimization can identify particular useful members of this family, a fundamental electromagnetic information theory should determine what can be achieved over the physically admissible family as a whole. Moreover, programmability can play a more fundamental role than selecting which transformation a system implements: information can itself be encoded into the tunable parameters. This possibility underlies backscatter communication in which data are mapped directly onto programmable scattering states, as well as wave-domain physical neural networks in which input data are encoded into tunable scattering properties and processed through the resulting wave--matter interactions. Programmability can therefore become intrinsic to how information is extracted, transferred and processed.

A fundamental electromagnetic information theory of programmable wave systems must address limits at two levels. At the first, system-specific level, the geometry, materials, losses and available controls of a given wave system are fixed, and one seeks the best sensing precision, information-transfer rate and computational performance that the resulting platform can achieve. The specified system may be modeled analytically or numerically, or realized experimentally. Such bounds establish the ultimate capabilities of a particular system independently of the optimization strategy used to configure it. At the second level, the system itself becomes variable: one asks what non-programmable and programmable physical resources bound these system-specific limits. Given prescribed resource budgets and physical constraints---such as aperture, volume, thickness, material response, loss, and the number and tuning range of independently controllable parameters---what is the greatest task performance that any physically admissible architecture could attain? Equivalently, what minimum resources are required to realize a prescribed functionality or level of performance? The first level establishes how closely a particular system approaches its own fundamental limits; the second provides principles for designing systems whose system-specific limits are fundamentally higher.

In this Perspective, we examine the emerging foundations of an electromagnetic information theory for programmable metasurfaces, and programmable wave systems more generally, from these two viewpoints. 
We first distinguish how information can enter a wave system through its excitation, its scattering structure or both, and how programmability allows the resulting input--output mapping to be controlled separately from how information is encoded. We then consider how fundamental limits developed for fixed wave systems can be extended to programmable systems. We next examine the non-programmable and programmable physical resources that determine the flexibility and expressivity of such systems, including thickness, nonlocality, mutual coupling, multiple scattering and the number of tunable degrees of freedom. Finally, we discuss further extensions involving the interface between wave-domain and digital processing, partially coherent and periodically time-modulated systems, and adaptive programmability.

\section{Information encoding and control in wave systems}
\label{sec:encoding}

\begin{figure*}[t]
    \centering
    \includegraphics[width=0.6\textwidth]{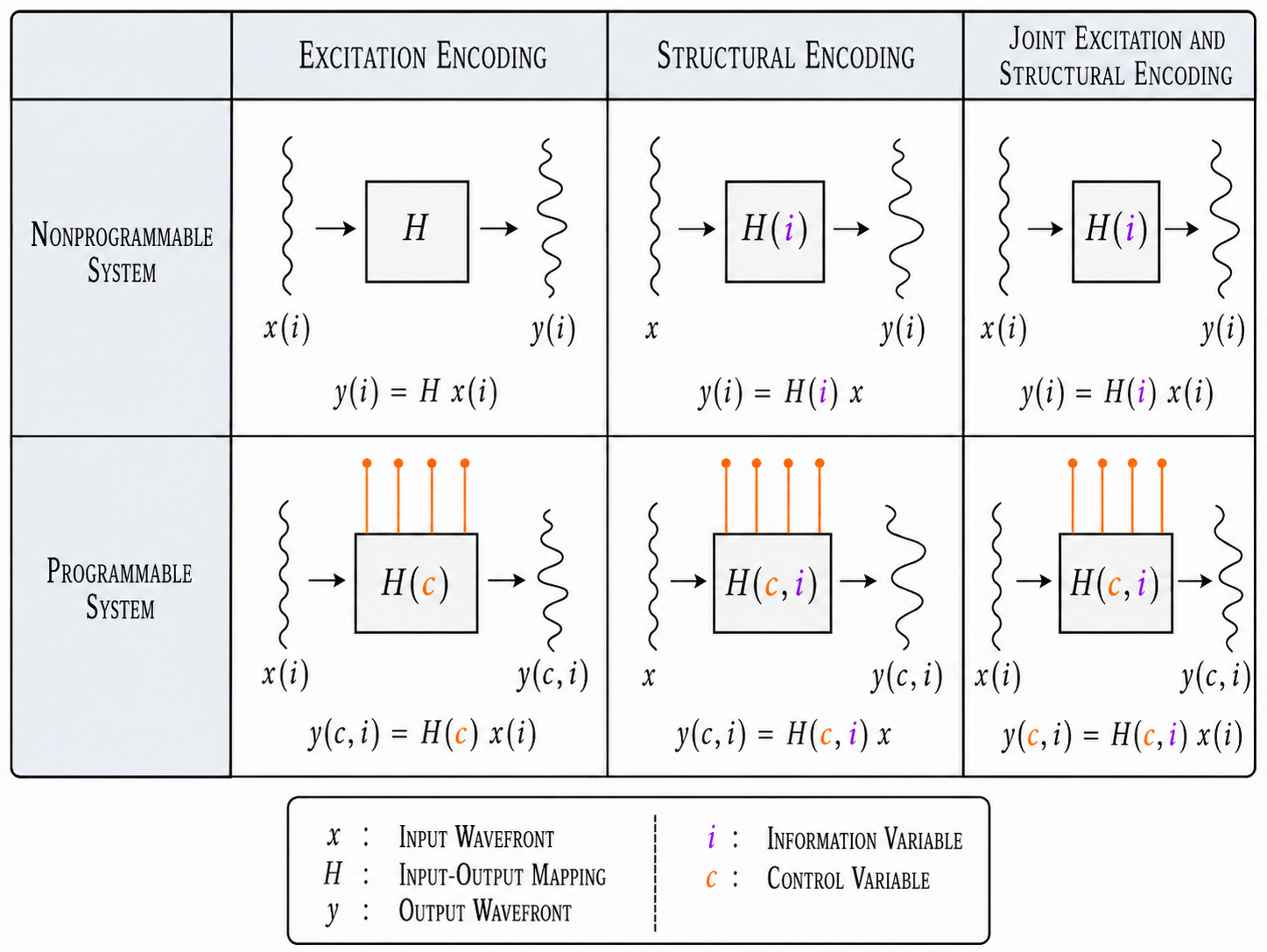}
    \caption{\textbf{Information encoding and control in wave systems.}
    The information variable $i$ can enter the electromagnetic problem through the input wavefront $x$ (excitation encoding), through the physical scattering structure represented by the input--output mapping $H$ (structural encoding), or through both (joint encoding). A programmable system provides an additional control variable $c$ that can modify the input--output mapping independently of the information variable. ``Non-programmable'' therefore denotes the absence of such an independent control variable; the mapping may nevertheless depend on $i$, for example when an unknown scene carries the information to be sensed. The variables $i$ and $c$ denote functional roles and need not correspond to physically separate degrees of freedom.}
    \label{fig:encoding}
\end{figure*}

Before asking what a programmable wave system can fundamentally achieve, it is useful to distinguish where the information enters the electromagnetic problem from how the wave system itself is controlled. Unless stated otherwise, we consider coherently excited passive systems whose electromagnetic response is linear and time invariant for each fixed combination of information-bearing state and control state\footnote{The assumed linearity and time invariance of the field response do not require coherent illumination.}; these states may nevertheless change between measurements, communication symbols or computational inputs. We denote by $i$ the information variable to be sensed, transferred or processed; by $x$ and $y$ the input and output wavefronts; and by $H$ the physical input--output mapping (Fig.~\ref{fig:encoding}). Information can enter through the excitation, $x(i)$, through the scattering structure, $H(i)$, or jointly through both. Programmability introduces an additional control variable $c$ over the scattering structure, giving mappings such as $H(c)$ or $H(c,i)$. In Fig.~\ref{fig:encoding}, ``non-programmable'' therefore means that no control variable $c$ is available independently of the information variable. It does not imply that $H$ must be independent of $i$: an unknown or information-bearing structure can itself modify the mapping. Conversely, $i$ and $c$ describe functional roles rather than necessarily distinct pieces of hardware; tunable degrees of freedom may carry information, provide control, or do both.

In communication systems, the conventional case is excitation encoding. A message $i$ is mapped to a transmitted wavefront $x(i)$, which subsequently propagates through the input--output mapping $H$. A programmable wave system can reshape that mapping through $c$, yielding $y(c,i)=H(c)x(i)$ while the information remains encoded in the excitation. Examples include reconfigurable intelligent surfaces (RISs) in smart radio environments~\cite{subrt2012intelligent,liaskos2018new,direnzo2019smart,direnzo2020smart}, dynamic metasurface antennas~\cite{shlezinger2021dynamic}, and, at optical frequencies, programmable multi-plane light converters~\cite{morizur2010programmable}. Structural encoding uses the same hardware with a different logic: the information changes the scattering properties themselves. This principle underlies backscatter communication---from passive eavesdropping devices of the 1940s to modern RFID---in which information is encoded by modulating a load or scattering state rather than actively generating an information-bearing carrier~\cite{brooker2013bug,stockman1948communication,landt2005}. Programmable metasurfaces enable extensions of this principle to many controllable scattering degrees of freedom, as in massive backscatter communication~\cite{li2021singlerf,zhao2020metasurface}. Excitation and structural encoding can also be combined, so that information is carried jointly by the excitation $x(i)$ and the input--output mapping $H(i)$~\cite{karasik2021adaptive}.

For sensing, the same classification applies, although the information-bearing state is generally imposed by the source or scene; the receiver must infer it from the measured response. When the quantity of interest is carried by an impinging field---for example, the direction of arrival of an unknown source---the information enters through excitation encoding, $x(i)$. In active sensing and imaging, by contrast, a known probe interrogates an unknown object or scene whose geometry, material properties or reflectivity determine how the field is scattered. The information then enters through structural encoding, $H(i)$, and the measured wavefront is $y(i)=H(i)x$. Structural encoding is therefore not necessarily deliberately engineered; it can be imposed by the physical scene that is being sensed. Programmability adds independent control over how this information is interrogated or collected, leading to $H(c,i)$. Joint encoding arises when both the impinging field and the scattering structure contain information to be sensed.

Wave-domain computation provides a third interpretation. In the conventional setting, data $i$ are encoded into an input wavefront $x(i)$, and a non-programmable or programmable physical operator implements $y(i)=Hx(i)$ or $y(c,i)=H(c)x(i)$. Structural encoding instead writes the computational input into tunable scattering properties, so that the physical input--output mapping itself depends on the data. This approach has recently been exploited in wave-based physical neural networks, in which input data are encoded into programmable structural states~\cite{eliezer2023tunable,momeni2023backpropagation,xia2024nonlinear,yildirim2024nonlinear,wanjura2024fully,li2024nonlinearencoding,rahman2025massivelyparallel,hammami2026expressivity}. Structural encoding is particularly consequential when $H(i)$ is a nonlinear function, because this makes the resulting information-to-output mapping nonlinear and thereby expands the class of functions that can be realized using the wave system. Such nonlinearity has two sources: the response of each tunable element can depend nonlinearly on its physical control (a varactor's reflection coefficient on its bias voltage, for example), and coupling between tunable elements makes the response nonlinear in the element states themselves (Box~1)~\cite{hammami2026expressivity}. More generally, joint excitation and structural encoding of the computational inputs could be explored.

Altogether, the taxonomy in Fig.~\ref{fig:encoding} thus separates two effects of programmability that are easily conflated. \textit{First,} the control variable $c$ creates a controllable family of electromagnetic mappings. \textit{Second,} programmable structural degrees of freedom can themselves become information-bearing. A comprehensive electromagnetic information theory for programmable metasurfaces should accommodate both effects across information extraction, transfer and processing. Because the same description applies to sensing, communication and computation, it also provides a natural language for systems that jointly sense, communicate and compute. In information-theoretic terms, $c$ is a channel state, i.e., a variable on which the conditional probability density of the output wavefront $y$ given the input wavefront $x$ depends, and so is $i$ under structural encoding. Excitation encoding gives $c$ control of the channel to be used, structural encoding uses the state itself as the message, and joint encoding combines the two. Channels with action-dependent states provide a related information-theoretic framework~\cite{weissman2010capacity}.

\section{System-specific limits of programmable wave systems}
\label{sec:prototype_specific}

\begin{figure*}[t]
    \centering
    \includegraphics[width=\textwidth]{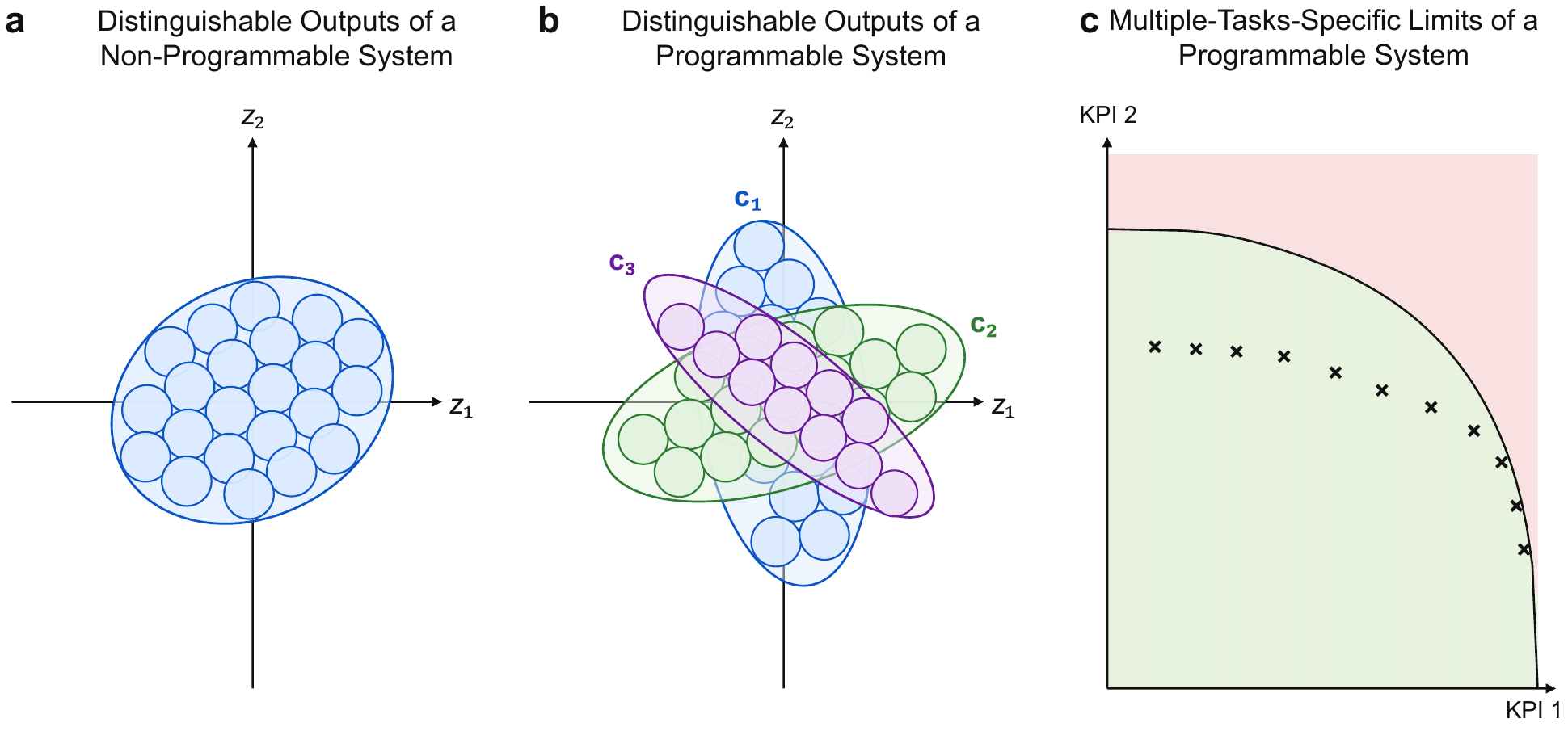}
    \caption{\textbf{System-specific limits of programmable wave systems.}
    \textbf{a,} In a non-programmable system, a linear operator maps energy-constrained inputs onto an output ellipsoid whose geometry is determined by its singular spectrum; non-overlapping balls schematically represent distinguishable output states (for Shannon capacity, the packing is over codewords spanning many channel uses).
    \textbf{b,} Different control states generate a correlated family of accessible output sets, illustrated here in a schematic two-dimensional output space. Their union gives the outputs accessible across all control states; overlap indicates outputs accessible under multiple control configurations.
    \textbf{c,} The green shaded region contains the achievable region in the space of two key performance indicators (KPIs), such as information-transfer rate and sensing precision in integrated sensing and communication, or target-mode strength and fidelity in output wavefront synthesis, with larger values indicating better performance. A small gap between the fundamental bound (solid line) and an achievable optimization outcome (cross symbols) certifies both near-optimality and bound tightness; a large gap does not distinguish optimization suboptimality from bound looseness.}
    \label{fig:prototype_specific}
\end{figure*}

A central concept for understanding system-specific limits is the distinguishability of the system's possible outputs. Practical detectors are limited by finite resolution or noise, such that only a finite number of output states can be distinguished. Under an input-energy constraint, a fixed linear operator maps the admissible inputs onto an ellipsoid whose semiaxis lengths in the output singular-vector basis are set by the singular values. This singular-value picture underlies electromagnetic degrees of freedom and communication-mode analyses: the effective electromagnetic degrees of freedom (EEMDOFs) quantify the effective number of independent field modes that remain significant relative to a relevant noise or resolution scale~\cite{miller2000communicating,piestun2000degrees,franceschetti2015landau,franceschetti2015scattered,miller2019waves,gustafsson2025degrees,zhu2024eit,wang2026continuous}. These modes determine the effective dimensionality available for carrying distinguishable information.\footnote{For a fixed input--output mapping $H$ with additive white Gaussian noise of power $N_0$ per mode and total input power $P$, the capacity is the waterfilling sum $\sum_k\log_2(1+p_k\sigma_k^2/N_0)$ over modes with singular values $\sigma_k$, maximized subject to $\sum_k p_k\le P$~\cite{telatar1999capacity}.} Distinguishability within that space can then be defined from a probabilistic (Shannon) or deterministic (Kolmogorov) perspective~\cite{Shannon1948,kolmogorov1959epsilon,franceschetti2017wave,lim2017information,wei2026electromagnetic}. In the former, reliably distinguishable codewords are separated by noise-limited uncertainty regions; in the latter, resolvable field states are separated by a prescribed resolution. Both therefore admit a geometric interpretation as packing problems within the accessible output set~\cite{migliore2008electromagnetics}, with distinguishability regions represented schematically in Fig.~\ref{fig:prototype_specific}a by non-overlapping spheres.

The relevant EEMDOFs depend on how information enters the system. For excitation encoding, they follow from the singular-value spectrum of the linear excitation-to-output operator. For structural encoding, the information-to-output mapping may instead be nonlinear, for example because coupling between tunable elements makes the electromagnetic response a nonlinear function of the information-bearing structural parameters. The EEMDOFs in the vicinity of a reference configuration can nevertheless be characterized from the Jacobian with respect to the tunable parameters~\cite{delhougne2026backscatter}. These local EEMDOFs can vary across reference configurations, and their distribution can be shaped through the system's excitation~\cite{delhougne2026backscatter}. When information-bearing parameters are to be estimated rather than decoded, the same Jacobian is the core of the Fisher information, whose inverse lower-bounds the error covariance of any unbiased estimator (the Cram\'er--Rao bound)~\cite{kay1993estimation}. System-specific sensing limits then take the form of optimal experiment design~\cite{pukelsheim2006optimal}, with the programmable metasurface as the experiment: choosing one or several admissible configurations to maximize a scalar measure of the combined Fisher information~\cite{han2026angular,wijekoon2026joint}.

Programmability extends the picture from the single operator $H$ (Fig.~\ref{fig:prototype_specific}a) to the complete physically admissible family $\mathcal{H}$ introduced above (Fig.~\ref{fig:prototype_specific}b). Different control states generate different accessible output sets, which are correlated because they originate from the same geometry, materials and control hardware. Their shapes, orientations, extents and degree of overlap depend on the control configuration. Depending on this overlap, some output states may be reachable under multiple control configurations, whereas other portions of the accessible sets may be configuration-specific. Figure~\ref{fig:prototype_specific}b schematically illustrates this family in a two-dimensional output space. Each fixed control state therefore has its own distinguishability packing, while programmability generates an ensemble of correlated packings. Which generalization of these packing and counting notions is relevant depends on how the control state is used. If $c$ is chosen independently of the message, held fixed during transmission and known to the transmitter and receiver, the best achievable capacity under a common input-power constraint is that of the best member of the family. If $c$ varies within a codeword, or itself carries information as in structural encoding, the system is a channel with state, for which information theory provides capacity formulas~\cite{shannon1958channels,gelfand1980coding,weissman2010capacity}. With the same input-power constraint imposed separately on each control configuration, the gain over the best single configuration is then bounded by the information that the control sequence itself conveys: at most $\log_2|\mathcal{C}|$ bits per reconfiguration for a finite set $\mathcal{C}$ of admissible control states, and, for analog controls, a noise-limited quantity governed by how distinguishable the configurations are through their overlapping output sets (Fig.~\ref{fig:prototype_specific}b). A key open question is how these functionals can be maximized, or at least bounded, in full-wave settings, where the correlations among the operators $H(c)$ and their nonlinear dependence on $c$ are fixed by Maxwell's equations and the hardware (Box~1).

Beyond characterizing the number of distinguishable outputs of a given programmable wave system, it is often important to understand the best performance that the system can achieve in a specific task. Optimizing the tunable configuration is typically a nonlinear, non-convex and high-dimensional inverse problem, so the computational cost of identifying a globally optimal configuration is often prohibitively large. Global performance bounds are therefore valuable because they can certify the near-optimality of an optimized solution or establish that a prescribed objective is unattainable. Such task-specific, system-specific limits concern the complete reachable family rather than any particular optimized configuration. For a single task metric, the limit is the best performance attainable over all admissible controls and any other available optimization variables. With several competing metrics, the corresponding object is an achievable performance region whose outer Pareto boundary represents the system-specific limit (Fig.~\ref{fig:prototype_specific}c). Optimization identifies particular achievable points within this region, whereas a fundamental bound constrains its outer boundary independently of the algorithm used to search for suitable configurations.

For a specified wave system, the family of possible outputs is shaped by the admissible continuous or discrete control states, static scattering contributions from non-tunable parts, and loss in both static and tunable parts. In addition, coupling effects intertwine the influence of the tunable elements on the output. An efficient way to incorporate these practically important aspects into a system-aware bound is to use a compact physics-based model. For lumped-element-programmable microwave systems, multiport-network theory provides a tractable route (Box~1). The mathematical structure of the model and its parameters, specified theoretically, obtained numerically or estimated from measurements, then capture these system-specific constraints and correlations. Several recent limits exploit cases in which the resulting optimization can be formulated as a quadratically constrained quadratic program and relaxed to a convex semidefinite program, whose global solution provides a certified bound on the original non-convex problem~\cite{salmi2025optimization,salmi2026bounds,delhougne2026mimo,del2026prototype}. Closely related formulations underlie earlier bounds on non-programmable or globally tunable nanophotonic devices~\cite{Gustafsson2020UpperBounds,molesky2020hierarchical,kuang2020computational,shim2024fundamental}: the underlying design problem of optimizing a pixelized dielectric distribution within a prescribed geometry parallels that of optimizing tunable loads within an otherwise fixed electromagnetic setup. Nonetheless, system-specific, physics-aware limits on information extraction, transfer and processing remain largely undeveloped, including estimation precision, MIMO channel capacity and computational expressivity. The first two have canonical information-theoretic functionals, the Fisher information and the channel capacity, but wave-domain computing still lacks an agreed measure of performance. Candidates include source-coding formulations for computing functions of the inputs~\cite{orlitsky2001coding}, the information-processing capacity of physical dynamical systems~\cite{dambre2012information}, and expressivity measures~\cite{hammami2026expressivity}; understanding when to choose which functional remains fertile open territory.

\begin{figure*}[tp]
\centering
\setlength{\fboxsep}{10pt}
\setlength{\fboxrule}{0.7pt}
\fbox{\begin{minipage}{\dimexpr\textwidth-2\fboxsep-2\fboxrule\relax}
\small
\setlength{\parindent}{0pt}
\setlength{\parskip}{0.5em}

\textbf{Box 1 | Compact physics-based models of programmable linear wave systems}

Multiport-network theory (MNT) provides a convenient abstraction between a full Maxwell description and the compact models required for system-aware analysis and optimization~\cite{direnzo2024electromagnetic}. It applies naturally to radio-frequency and microwave programmable metasurfaces whose reconfiguration is implemented by lumped elements such as PIN diodes or varactors. Each tunable lumped element is represented as a tunable load terminating a ``virtual'' port. The system is then partitioned into a \textit{static subsystem}, containing the electromagnetic structure, and a \textit{tunable subsystem}, containing the tunable loads. The two are interconnected through the ``virtual'' ports. Each subsystem is described by a multiport scattering matrix. When the tunable subsystem consists of an ensemble of individual tunable loads, its scattering matrix is diagonal; for beyond-diagonal tunable load networks~\cite{shen2022modeling}, tunable interconnections between ``virtual'' ports can instead make this matrix non-diagonal. Yet, envisioned implementations of these beyond-diagonal networks typically rely on tunable lumped elements, so the tunable subsystem is still an ensemble of individual tunable loads characterized by a diagonal scattering matrix, with the ``virtual'' ports placed at the individual loads and the static interconnections included in the static subsystem~\cite{delhougne2025diagonal}. Connecting the two subsystems through the ``virtual'' ports yields the following end-to-end mapping between the external input and output ports:
\[
H(c)=H_0+A\left[I-\Phi(c)C\right]^{-1}\Phi(c)B .
\]
Here, $\Phi(c)$ is the scattering matrix of the tunable subsystem, $H_0$ represents the static contribution to the input--output response that does not involve scattering through the tunable loads, $A$ and $B$ couple the ``virtual'' ports to the external output and input ports, respectively, and $C$ describes coupling among the virtual ports. The inverse term captures repeated interactions between tunable elements and hence coupling-induced multiple scattering. Quantization, state-dependent loss and other hardware restrictions enter through the admissible values of $\Phi(c)$.

For experimental prototypes, the required MNT parameters are rarely known sufficiently accurately from design information alone because of fabrication inaccuracies and incomplete or proprietary hardware information. Directly measuring the underlying multiport scattering matrices is generally impractical because the ``virtual'' ports are not connectorized and typically greatly outnumber the ports of conventional measurement equipment. Instead, proxy MNT parameters can be inferred from end-to-end measurements over multiple control configurations~\cite{delhougne2025experimental}. These internal parameters need not be uniquely identifiable; such ambiguities are immaterial provided that the resulting proxy model accurately predicts the observable scattering response over the admissible control configurations. The resulting freedom to choose among equivalent parameterizations may itself be exploitable~\cite{delhougne2026mobius}, but this possibility remains largely unexplored. Residual model error should be accounted for by bounds valid for all models consistent with the measurements; a communication-specific example is the max--min capacity of a compound channel, when one code must work across the uncertainty set~\cite{blackwell1959capacity}.

MNT is one of several reduced-order models with this structure. At optical and terahertz frequencies, where tuning shifts resonant modes rather than lumped loads, temporal coupled-mode theory (CMT) is the standard formulation~\cite{haus1984waves,fan2003temporal,suh2004temporal}. Its scattering matrix, $S(c)=S_d+D\left[i(\omega-\Omega(c))+\Gamma\right]^{-1}K^{\mathsf{T}}$, can be recast exactly in the form above by separating the control-dependent detuning from a fixed reference response. Coupled-dipole models, also used for dynamic metasurface antennas and RIS-parametrized channels, have the same form~\cite{faqiri2023physfad}, with polarizabilities in place of $\Phi$ and the dipole--dipole interaction matrix in place of $C$; so, in the continuum limit, does the volume-integral T-operator $(\chi^{-1}-G_0)^{-1}$, with susceptibility $\chi$ and background Green's function $G_0$.
As with MNT proxies, CMT parameters fitted to measured spectra are not unique, because a change of modal basis leaves $S$ unchanged.
\end{minipage}}
\end{figure*}

Programmable interferometric meshes provide an instructive idealized benchmark that largely abstracts away from the hardware constraints considered above and for which the reachable family is unusually well understood. Appropriately interconnected tunable interferometers can realize arbitrary unitary transformations between a finite set of modes and, with additional control of the singular values, more general linear transformations subject to the physical constraints of the system~\cite{miller2013selfconfiguring,bogaerts2020programmable}. Moreover, some useful transformations can be accessed without solving a global inverse problem: suitably designed self-configuring architectures can automatically establish optimal orthogonal communication channels using local measurements and feedback~\cite{miller2013optimalchannels,annoni2017unscrambling,seyedinnavadeh2024determining}. Self-configuring meshes can also measure incoming fields, with the measured field components encoded in the configured settings and output powers~\cite{miller2013selfconfiguring}. Related progressive design and self-configuration principles have been proposed for forward-only multilayer metastructures~\cite{miller2026moving}.

\section{Physical resources that bound system-specific limits}
\label{sec:resources}

\begin{figure*}[t]
    \centering
    \includegraphics[width=0.5\textwidth]{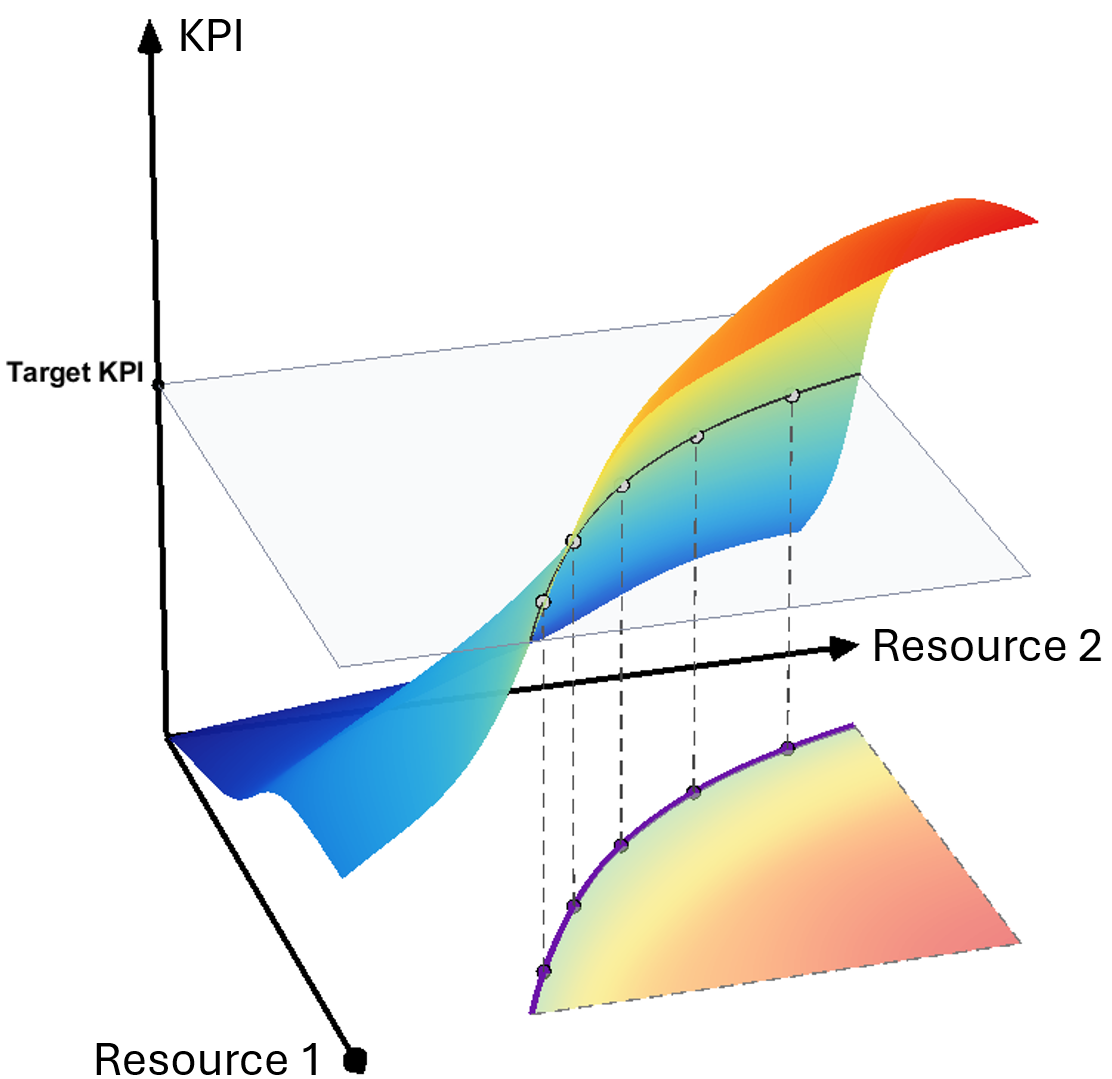}
    \caption{\textbf{Resource bounds on system-specific performance.}
    The system-specific limit of a programmable wave system generally depends on a multidimensional vector of budgets for non-programmable and programmable physical resources; two representative dimensions are shown schematically. KPI denotes a key performance indicator, with larger values indicating better performance. The surface represents the greatest task performance attainable by any physically admissible architecture for each resource combination. For a prescribed target performance, multiple resource combinations may suffice; their lower boundary identifies the Pareto-minimal resource combinations (solid purple line).}
    \label{fig:resources}
\end{figure*}

System-specific limits constrain the performance of a particular programmable wave system. Understanding how these limits depend on architecture and resources requires moving from the control-dependent family of input--output operators $\mathcal{H}=\{H(c):c\in\mathcal{C}\}$ of a fixed system to the non-programmable and programmable physical resources required to support such a family. The system must now be allowed to vary, subject to prescribed budgets for non-programmable and programmable physical resources and physical constraints.
The resources are intrinsically multidimensional and can include aperture, volume and thickness; available material response and bandwidth; the number, resolution and tuning range of programmable degrees of freedom; the number of radio-frequency (RF) chains that feed or read out the system; the rate at which the programmable degrees of freedom can be reconfigured; the overhead of acquiring the channel information needed to configure them; and electromagnetic interactions such as coupling and multiple scattering. Loss and other constraints determine how effectively these resources can be converted into task performance. For a given resource vector and set of physical constraints, one can ask for the greatest system-specific task-performance limit attainable by any physically admissible architecture. Conversely, for a prescribed target performance, one can ask which resource combinations are fundamentally required. Figure~\ref{fig:resources} illustrates both viewpoints for two representative resource dimensions: the surface gives the greatest task performance attainable for each resource combination, while a prescribed target performance defines a Pareto boundary of resource combinations sufficient to attain it.

Spatial extent and material properties provide fundamental electromagnetic resources even before programmability is considered. Existing electromagnetic bounds for non-programmable and globally tunable nanophotonic devices connect quantities such as volume, material response and loss to achievable scattering, absorption and field concentration~\cite{Miller2026FundamentalLimits,chao2022physical,Gustafsson2020UpperBounds,kuang2020computational,shim2024fundamental}, and to achievable input--output transformations~\cite{molesky2022t}. These formulations allow the material distribution to vary within a pixelized representation of a design region whose shape, extent and available material responses are specified. By contrast, overlapping nonlocality provides an architecture-agnostic channel-count bound derived directly from the desired input--output transformation. It lower-bounds the number of independent channels that must cross a cut---an imaginary dividing surface that partitions both the input and output regions into two sides. This channel-count requirement can translate into a minimum-thickness bound. For each cut, we isolate the two parts of the input--output matrix that map inputs on one side to outputs on the other. In the usual case of a reciprocal system, we count their significant singular values, add the counts for both directions, and take the largest total over all cuts~\cite{miller2023thickness}. This channel-count bound applies whether or not the channels themselves are programmable. For propagation in a homogeneous medium, diffraction-limited channel packing imposes a minimum transverse area and, for a prescribed lateral extent, a minimum device thickness~\cite{miller2023thickness}. Guided interconnections, including those envisioned for so-called ``beyond-diagonal'' tunable load networks~\cite{shen2022modeling}, may allow the required channels to be routed within a plane. Establishing corresponding minimum-thickness bounds for such architectures remains an open problem. Relatedly, the spatial complexity of a desired operation constrains how compactly it can be implemented in the wave domain~\cite{li2025spatial}. The available volume also constrains how many wave channels can couple strongly to the exterior. Tunneling arguments provide a size-dependent bound on the number of such channels~\cite{miller2025tunnelling}, while shape-independent bounds constrain their coupling strengths and associated information capacities between finite domains~\cite{kuang2025coupling}. Although most of these results concern fixed wave systems, the same physical constraints necessarily apply to every operator $H(c)$ accessible to a programmable system.

Programmability introduces additional resources. Increasing the number of independently tunable parameters enlarges the control space $\mathcal{C}$, but it does not necessarily enlarge $\mathcal{H}$ by a corresponding number of independent electromagnetic transformations. Different controls can produce redundant changes of the input--output operator, whereas suitable placement and connectivity can make their effects complementary. Tuning range and resolution likewise determine how widely and densely the physically accessible set of operators can be traversed. Under idealized assumptions of lossless scattering and independent, effective and continuously tunable parameters, the minimum number of tuning parameters required for certain wave-routing tasks can already be predicted analytically~\cite{alhulaymi2025ccon}. Parameter counting also informs constructive designs: ideal interferometer meshes implement universal linear transformations with adjustable-parameter counts that match the dimensionality of the target operator~\cite{miller2013selfconfiguring,bogaerts2020programmable}. More generally, a central open question is how the number and nature of programmable degrees of freedom constrain the dimensionality, diversity and task-relevant variation of the accessible operator set $\mathcal{H}$. Temporal resources also matter. When information is encoded in the control state, as in backscatter communication, the achievable rate is at most the information per configuration multiplied by the reconfiguration rate, so switching speed constrains performance as directly as aperture or thickness. Moreover, acquiring the channel information needed for configuration can consume a substantial part of a coherence interval~\cite{hassibi2003training}; the overhead depends on prior knowledge and the available measurements.

Electromagnetic interactions among the tunable elements constitute another important design variable. Mutual coupling and multiple scattering were complications that had to be retained in the system-specific models of the preceding section; at the architecture level, however, they can also become resources. Stronger coupling can increase the number of significant interactions of the wave with the tunable components, thereby increasing the influence of the programmable degrees of freedom on the electromagnetic response~\cite{prodhomme2025friend,prodhomme2025benefits}. 
When information is encoded in structural parameters, repeated interactions with the tunable elements can make the output a nonlinear function of those parameters. Increasing the number of scattering events can strengthen this structural nonlinearity and thereby increase the expressivity of wave-based transformations~\cite{eliezer2023tunable,momeni2023backpropagation,xia2024nonlinear,yildirim2024nonlinear,wanjura2024fully,hammami2026expressivity}.
Yet, this does not imply that stronger coupling or more multiple scattering is universally beneficial. General linear transformations, subject to passivity, can also be implemented in feedforward interferometric meshes without recurrent scattering~\cite{miller2013selfconfiguring}. The relevant question is how these interactions reshape the task-relevant variations within $\mathcal{H}$ while simultaneously affecting loss, bandwidth, sensitivity and robustness.

The different resources are generally neither independent nor interchangeable in a simple manner. Increasing thickness can enable greater spatial nonlocality; adding tunable elements can alter the coupling topology; and exploiting repeated interactions can enhance control while simultaneously increasing dissipation or resonant sensitivity. These dependencies also suggest distinct architectural routes for enlarging the task-relevant variation within $\mathcal{H}$. One route is to exploit stronger electromagnetic interactions, with associated sensitivity and robustness trade-offs~\cite{prodhomme2025friend,prodhomme2025benefits}. Another is to increase the interaction depth by stacking multiple programmable metasurface layers, as in stacked intelligent metasurfaces, at the expense of greater thickness and typically more tunable components~\cite{an2023stacked,SIMsurvey,miller2026moving}. Consequently, raising system-specific limits is intrinsically a multi-resource design problem rather than a matter of maximizing any single physical parameter. At a fixed resource vector, different architectures may attain different system-specific limits, while a fundamental resource--performance envelope would bound them all. This ordering holds for the true limits but not necessarily for computed bounds: for instance, a relaxation-based system-aware bound can be looser than an architecture-independent one, as recently observed for the shaping of partially coherent microwaves~\cite{delhougne2026partial}, so the tightest certificate for a given system may come from the architecture level. In the reverse direction, a target performance will generally not correspond to a unique minimum resource vector: different combinations of resources such as physical extent, programmable degrees of freedom and electromagnetic interactions may achieve the same target. Unless these resources are assigned a common scalar cost, the appropriate notion of minimum resources is therefore a Pareto boundary of resource combinations.

The broader challenge is to bound the best task performance attainable across architectures sharing prescribed budgets for non-programmable and programmable physical resources, and to identify architectures and resource combinations that enable performance close to these bounds. This requires accounting for the full accessible operator family $\mathcal{H}$ of each architecture. Such relations would identify the resource bottlenecks for a desired sensing, communication or computing task, when additional programmable parameters cease to provide useful control, and when greater physical extent or stronger electromagnetic interactions can compensate for fewer tunable elements. They would ultimately connect the non-programmable and programmable physical resources of an architecture to the set of operators $\mathcal{H}$ that it can realize, and from there to the system-specific information extraction, transfer or processing performance that can be achieved.

\section{Outlook}
\label{sec:outlook}

The central challenge is to turn the conceptual framework outlined here into quantitative and experimentally testable limits. At the system-specific level, fundamental bounds remain to be established for many practically relevant measures of information extraction, transfer and processing, and more generally for the distinguishability and task-relevant diversity supported by the complete programmable family $\mathcal{H}$. At the architecture level, an even broader open problem is to determine how non-programmable and programmable physical resources constrain these system-specific limits, and which combinations of resources are fundamentally required to reach a prescribed level of performance. Progress on both levels will require theoretical tools that retain the relevant electromagnetic physics while remaining computationally tractable and applicable to realistic systems. Experimentally calibrated physics-based models provide one route to identifying such bounds for fabricated hardware prototypes and determining how closely actual programmable wave systems approach their fundamental limits. Box~2 lists concrete open problems at both levels.

\begin{figure*}[tp]
\centering
\setlength{\fboxsep}{10pt}
\setlength{\fboxrule}{0.7pt}
\fbox{\begin{minipage}{\dimexpr\textwidth-2\fboxsep-2\fboxrule\relax}

\small
\setlength{\parindent}{0pt}
\setlength{\parskip}{0.5em}

\textbf{Box 2 | Open problems}

\begin{enumerate}[leftmargin=1.7em,itemsep=0pt,parsep=0.5em,topsep=0pt,partopsep=0pt]
\item \textit{Performance measures for wave-domain computing:} establish task-relevant functionals that support meaningful bounds and comparisons between programmable wave systems.
\item \textit{Information-encoding strategies:} quantify how excitation encoding, structural encoding and their combination affect attainable sensing, communication and computing performance, and identify when each is advantageous under common resource budgets.
\item \textit{System-specific limits:} bound sensing precision, communication rates and computing performance over the physically admissible configurations and excitations of a given system, including joint encoding in the excitation and control state.
\item \textit{Architecture-agnostic resource--performance envelopes:} bound the best task performance attainable across architectures sharing prescribed budgets for non-programmable and programmable physical resources, including reconfiguration time and channel-state acquisition overhead, and identify resource combinations required for a target performance.
\end{enumerate}
\end{minipage}}
\end{figure*}

Several further extensions are conceivable for an even more comprehensive electromagnetic information theory of programmable wave systems. 
\textit{First,} loss of task-relevant information at the interface between the wave and digital domains should be carefully accounted for and minimized, for example through end-to-end optimization that includes the programmable electromagnetic response, analog-to-digital conversion and downstream digital processing~\cite{wang2021dynamic,delhougne2026adc}.
\textit{Second,} the coherent, time-invariant setting considered so far should be generalized to partially coherent fields and intensity-only measurements, for which fundamental bounds and programmable approaches to measurement and processing are beginning to emerge~\cite{zhang2019concentration,roques2024measuring,Miller2026FundamentalLimits,delhougne2026partial,kienesberger2026metaimagers,yazdnian2026wireless}, and to time-modulated tunable elements, which open additional harmonic-frequency dimensions~\cite{zhang2018spacetime,viikari_timeFloquet}.
\textit{Third,} programmability can be adaptive: measurements acquired under one control state can inform subsequent control states. The corresponding fundamental limits should then extend from individual configurations to control policies that acquire and manipulate information sequentially. Sequential experiment design provides a mature framework for such policies, choosing subsequent configurations according to a task-relevant statistical criterion informed by past measurements~\cite{chernoff1959sequential}. Together with progress on the open questions outlined in Box~2, these extensions could help establish a general theory connecting electromagnetic physics, programmable resources and information-processing performance.

\end{document}